\documentclass[a4paper,10pt]{article}

\usepackage{amsmath}
\usepackage{amsthm}
\usepackage{amssymb}
\usepackage{graphicx}
\usepackage[labelformat=simple]{subfig}
\usepackage{float}
\usepackage[nameinlink]{cleveref}
\usepackage{verbatim}
\usepackage{xcolor}
\usepackage{complexity}
\usepackage{dsfont}
\newclass{\threesat}{3SAT}
\usepackage[inline]{enumitem}
\usepackage{thm-restate}
\usepackage{todonotes}
\usepackage{microtype}

\crefname{conjecture}{Conjecture}{Conjectures}
\crefname{proposition}{Proposition}{Propositions}
\crefname{lemma}{Lemma}{Lemmata}
\crefname{theorem}{Theorem}{Theorems}
\crefname{section}{Section}{Sections}
\crefname{appendix}{Appendix}{Appendices}
\crefname{figure}{Fig.}{Figs.}
\Crefname{figure}{Figure}{Figures}

\newcommand{\pa}{\mathcal{P}}
\newcommand{\cater}{\mathcal{C}}
\newcommand{\tree}{\mathcal{T}}

\def\diam{\operatorname{diam}}
\def\d{\operatorname{dist}}
\def\O{\mathcal{O}}
\def\dm{\operatorname{d}_{\min}}
\def\dM{\operatorname{d}_{\max}}
\def\dcg{\operatorname{DCG}_{S}}

\newcommand{\Wlog}{W.l.o.g.}

\newcommand{\Conv}{\operatorname{Conv}}

\newcommand\blfootnote[1]{%
	\begingroup
	\renewcommand\thefootnote{}\footnote{#1}%
	\addtocounter{footnote}{-1}%
	\endgroup
}

\usepackage{authblk}
\usepackage{listings}
\begin{document}
\date{}
\author[1]{Oswin Aichholzer}%\orcidID{0000-0002-2364-0583}}
\author[1]{Julia Obmann}
\author[2]{Pavel Paták}
\author[3]{Daniel Perz}%\orcidID{0000-0002-6557-2355}}
\author[4]{Josef Tkadlec}%\orcidID{0000-0002-1097-9684}}
\author[1]{Birgit Vogtenhuber}%\orcidID{0000-0002-7166-4467}}

\affil[1]{Graz University of Technology, Institute of Software Technology}
\affil[2]{Czech Technical University in Prague, Department of Applied Mathematics}
\affil[3]{University of Perugia, Department of Engineering}
\affil[4]{Charles University Prague, Computer Science Institute}

\title{Disjoint Compatibility via graph classes\thanks{Research on this work was initiated at the 6th Austrian-Japanese-Mexican-Spanish Workshop on Discrete Geometry and continued during the 16th European Geometric Graph-Week, both held near Strobl, Austria.
		We are grateful to the participants for the inspiring atmosphere. We especially thank Alexander Pilz for bringing this class of problems to our attention.
		D.P. was partially supported by the FWF grant I 3340-N35 (Collaborative DACH project \emph{Arrangements and Drawings}).
		The research stay of P.P. at IST Austria was funded by the project CZ.02.2.69/0.0/0.0/17\_050/0008466 Improvement of internationalization in the field of research and development at Charles University, through the support of quality projects MSCA-IF.}}

\maketitle
%\graphicspath{{figures/}}

\begin{abstract}
Two plane drawings of graphs on the same set of points are called disjoint compatible if their union is plane and they do not have an edge in common.
		Let $S$ be a convex point set of $2n \geq 10$ points and let $\mathcal{H}$ be a family of plane drawings on $S$.
		Two plane perfect matchings $M_1$ and $M_2$ on $S$ (which do not need to be disjoint nor compatible) are \emph{disjoint $\mathcal{H}$-compatible} if there exists a drawing in $\mathcal{H}$ which is disjoint compatible to both $M_1$ and $M_2$.
		In this work, we consider the graph which has all plane perfect matchings as vertices and where two vertices are connected by an edge if the matchings are disjoint $\mathcal{H}$-compatible.
		We study the diameter of this graph when $\mathcal{H}$ is the family of all plane spanning trees, caterpillars or paths.
		We show that in the first two cases the graph is connected with constant and linear diameter, respectively, while in the third case it is disconnected.
		\blfootnote{\begin{minipage}[l]{0.28\textwidth} \includegraphics[trim=10cm 6cm 10cm 5cm,clip,scale=0.15]{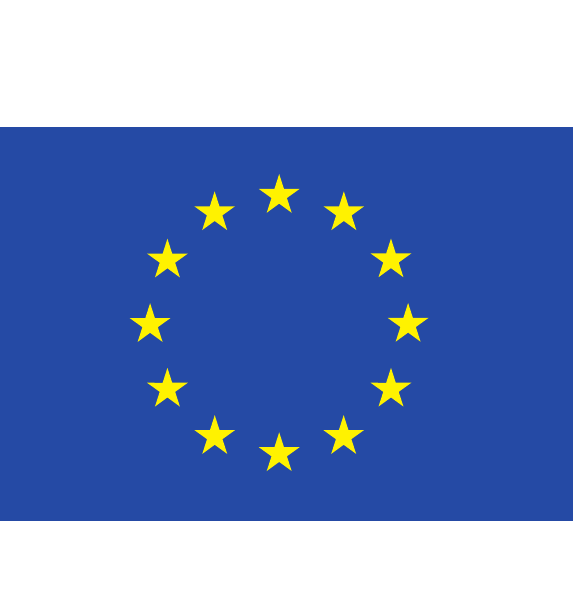} \end{minipage} \hspace{-2.0cm} \begin{minipage}[l][0,5cm]{0.83\textwidth}\vspace{-0.16cm}
				This project has received funding from the European Union's Horizon 2020 research and innovation programme under the Marie Sk\l{}odowska-Curie grant agreement No 734922.
		\end{minipage}}

\end{abstract}

\section{Introduction}
In this work we study straight-line drawings of graphs.
Two plane drawings of graphs on the same set $S$ of points are called \emph{compatible} if their union is plane. The drawings are \emph{disjoint compatible} if they are compatible and do not have an edge in common. For a fixed class $\cal{G}$, e.g.\ matchings, trees, etc., of plane geometric graphs  on $S$ the (disjoint) \emph{compatibility graph} of $S$ has the elements of $\cal{G}$ as the set of vertices and an edge between two elements of $\cal{G}$ if the two graphs are (disjoint) compatible. For example, it is well known that the (not necessarily disjoint) compatibility graph of plane perfect matchings is connected~\cite{hhn02,houle}.
Moreover, in~\cite{abdgh09} it is shown that there always exists a sequence of at most $\O(\log n)$ compatible (but
not necessarily disjoint) perfect matchings between any two plane perfect matchings of a set
of $2n$ points in general position, that is, the graph of perfect matchings is connected
with diameter~$\O(\log n)$.
On the other hand, Razen~\cite{Razen2008} provides an example of a point set where this diameter is $\Omega(\log n/ \log \log n)$.

Disjoint compatible (perfect) matchings have been investigated in~\cite{abdgh09} for sets of $2n$ points in general position. The authors showed that for odd $n$ there exist
perfect matchings which are isolated vertices in the disjoint compatibility graph and posed the following conjecture:
For every perfect matching with an even number of edges there exists a disjoint
compatible perfect matching. This conjecture was answered in
the positive by Ishaque, Souvaine and T{\'o}th~\cite{ishaque} and it was mentioned that for even $n$ it remains an open problem whether the disjoint compatibility graph is always connected.
In~\cite{aam15} it was shown that for sets of $2n \geq 6$ points in convex position
this disjoint compatibility graph is (always) disconnected.

Both concepts, compatibility and disjointness, are also used in combination with different geometric graphs.  For example, in~\cite{houle} it was shown that the flip-graph of all triangulations that admit a (compatible) perfect matching, is connected\footnote{In the flip-graph, two triangulations are connected if they differ by a single edge.}. It has also been shown that for every graph with an outerplanar embedding there exists a compatible plane perfect matching~\cite{aght}.  Considering plane trees and simple polygons, the same work provides bounds on the minimum number of edges a compatible plane perfect matching must have in common with the given graph. For simple polygons, it was shown in~\cite{pilz2020augmenting} that it is $\NP$-hard to decide whether there exist a perfect matching which is disjoint compatible to a given simple polygon. See also the survey~\cite{ht13} on the related concept of compatible graph augmentation.

In a similar spirit we can define a bipartite disjoint compatibility graph, where the two sides of the bipartition represent two different graph classes.
Let one side be all plane
perfect matchings of $S$ while the other side consists of all plane spanning trees of $S$.
Edges represent the pairs of matchings and trees which are disjoint compatible.
Considering connectivity of this bipartite graph, there trivially exist isolated
vertices on the tree side -- consider a spanning star, which can not have any disjoint compatible matching. Thus, the question remains whether there exists a bipartite connected subgraph which contains all vertices representing plane perfect matchings.

\begin{figure}[tb]
	\centering
	\includegraphics[page=1,scale=1]{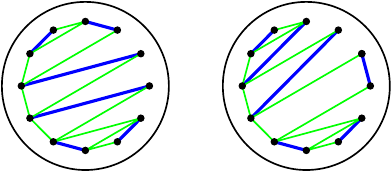}
	\caption{Two plane perfect matchings (in blue) on the same set of twelve points in convex position which are disjoint $\tree$-compatible. The complying disjoint compatible spanning tree is drawn in green.
	}
	\label{fig:tree-compatible}
\end{figure}

This point of view leads us to a new notion of adjacency for perfect matchings.
For a given set~$S$ of $2n$ points and a family $\mathcal{H}$ of drawings on $S$,
two plane perfect matchings $M_1$ and $M_2$ (which do not need to be disjoint nor compatible) are \emph{disjoint $\mathcal{H}$-compatible} if there exists a drawing $D$ in $\mathcal{H}$ which is disjoint compatible to both $M_1$ and $M_2$; see \cref{fig:tree-compatible} for an example.
The disjoint $\mathcal{H}$-compatibility graph $\dcg(\mathcal{H})$ has all plane perfect matchings of $S$ as vertices.
We have an edge between the vertices corresponding to $M_1$ and $M_2$ if $M_1$ and $M_2$ are disjoint $\mathcal{H}$-compatible.
In other words, they are two steps apart in the corresponding bipartite disjoint compatibility graph.
Rephrasing the above question, we ask whether $\dcg(\mathcal{H})$ is connected.
Recall that the disjoint compatibility graph for perfect matchings alone is not connected (see~\cite{aam15,abdgh09}).

In this work we study the case where $S$ is a set of $2n$ points in convex position and consider the cases where
$\mathcal{H}$ is the family $\tree$ of all plane spanning trees,
the family $\cater$ of all plane spanning caterpillars, or
the family $\pa$ of all plane spanning paths.
We show that $\dcg(\tree)$ and $\dcg(\cater)$ are connected if
$2n \geq 10$.
In that case the diameter of $\dcg(\tree)$ is either 4 or 5, independent of $n$,
and the diameter of $\dcg(\cater)$ is $\O(n)$.
For $n=2$, $\dcg(\tree)$ and $\dcg(\cater)$ are also connected.
While for $4\leq n \leq 10$, $\dcg(\tree)$ and $\dcg(\cater)$ are disconnected. This was verified by computer.
On the other hand we show that $\dcg(\pa)$ is disconnected.

From here on, if not said otherwise, all matchings, trees, caterpillars and paths are on point sets in convex position and are plane.
Hence, we omit the word 'plane' for these drawings.
Further, all matchings considered in this work are perfect matchings.
This work is partially based on the master's thesis of the second author~\cite{ObmannThesis}.

\section{Preliminaries}

Throughout this article let $S$ be a set of $2n$ points in the plane in convex position.
The edges of a drawing on $S$ can be classified in the following way.
	We call edges, that are spanned by two neighboring points of $S$, \emph{perimeter edges}; all other edges spanned by $S$ are called \emph{diagonals}.
	We call matchings without diagonals \emph{perimeter matchings}.
		Note that there are exactly two perfect perimeter matchings.
		We label the perimeter edges alternately even and odd.
		The \emph{even perimeter matching} consists of all even perimeter edges.
		The \emph{odd perimeter matching} consists of all odd perimeter edges.

\begin{figure}[tb]
	\centering
	\includegraphics[page=1,scale=1]{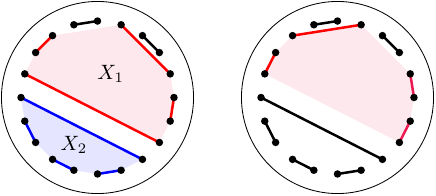}
	\caption{Left: A matching $M$ and two semicycles $X_1$ (red edges) and $X_2$ (blue edges) with their convex hulls. The cycle $\overline{X_1}$ is an inside $4$-cycle, since the boundary of the red shaded area contains at least two (in fact three) diagonals. The cycle $\overline{X_2}$ is a $4$-ear.
		Right: The matching resulting from rotating the cycle $X_1$.}
	\label{fig:rotation}
\end{figure}

	Looking at a matching $M$ on $S$, the edges of $M$ split the convex hull of $S$	into regions, such that no edge of $M$ crosses any region.
	More formally, we call a set $X \subset M$ of $k\geq 2$ matching edges a \emph{$k$-semicycle} if no edge of $M$ intersects the interior of the convex hull of $X$.
	Further, we call the boundary of the convex hull of $X$ a $k$-cycle, denoted by $\overline{X}$.
	If $\overline{X}$ contains at least two diagonals of $S$, then we call $X$ an \emph{inside $k$-semicycle}.
	Otherwise, we call $X$ a \emph{$k$-semiear} (this includes perimeter matchings); see \cref{fig:rotation}.
	Analogously, we denote cycles as \emph{inside $k$-cycles} or \emph{$k$-ears}, respectively.

	Consider a perfect matching $M$ and a semicycle $X$ of $M$. We say that we \emph{rotate} $X$ if we take all edges of $M$ and replace $X$ by $\overline{X}\backslash X$, which gives us a perfect matching $M'$.
	So the symmetric difference of $M$ and $M'$ is exactly $\overline{X}$.

\section{Disjoint compatibility via spanning trees}\label{sec:trees}
In this section we show that for convex point sets $S$ of $2n \geq 10$ points, the disjoint compatibility graph $\dcg(\tree)$ is connected.
We further prove that the diameter is upper bounded by $5$. The idea is that any matching on $S$ has small distance to one of the two perimeter matchings and those themselves are close to each other in $\dcg(\tree)$.
First we show that arbitrarily many inside cycles can be simultaneously rotated in one step.

\begin{restatable}{lemma}{lemmaInsideCycle}\label{insidecycle}
	Let $M$ and $M'$ be two matchings whose symmetric difference is a union of disjoint inside cycles. Then $M$ and $M'$ are $\tree$-compatible to each other.
\end{restatable}

\begin{figure}
	\centering
	\vspace{-3ex}
	\includegraphics[page=2,scale=0.95]{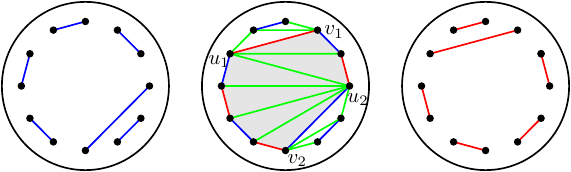}
	\caption{Two plane matchings (in blue and red) on $S$ which whose symmetric difference is an inside cycle. The complying disjoint compatible spanning tree is drawn in green.
	}
	\label{fig:tree-rotation}
\end{figure}

\begin{proof}
	First we focus on one inside cycle $C$.
	Let $u_1v_1$ and $u_2v_2$ be two diagonals of $\overline{X}$, labeled as in \cref{fig:tree-rotation}.
	Note that $v_1$ and $u_2$ could be the same point. % if each of $M$ and $M'$ contains one of $u_1v_1$ and $u_2v_2$.
	We take the edges from $u_1$ to any point between $v_1$ and $u_2$ and from $u_2$ to any point between $v_2$ and $u_1$ including $u_1$. 
	This yields a tree $T_1$ on the points of $X$ except $v_1$ and $v_2$.  
	
	If we have multiple disjoint inside cycles, we construct a tree $B'_j$ in this way for every inside cycle $C_j$.
	
	\begin{figure} %[tb]
		\centering
		\vspace{-3ex}
		\includegraphics[scale=1]{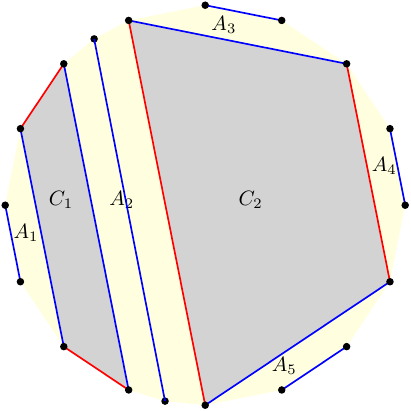}
		\caption{Two plane matchings (in blue and red) on $S$ which whose symmetric difference are two inside cycles ($C_1$ and $C_2$). The yellow regions are the convex hulls of the $A_i$s with $1\leq i \leq 5$.
		}
		\label{fig:inside-cycles-parts}
	\end{figure}
	
	Note that the inside cycles splits the convex hull of $S$ into multiple parts.
	We denote each part with $A_i \subset S$ where each $A_i$ is chosen maximal in the sense that it also contains the vertices of the bounding diagonals, see \cref{fig:inside-cycles-parts}.
	In other words, the $A_i$s are chosen such that the intersection of any inside cycle and any $A_i$ contains at most two vertices, any two distinct $A_i$ have at most a point in common
	and for any diagonal of an inside cycle, there exists exactly one index $i$ such that $A_i$ contains the vertices of the diagonal.
	Further, let $M_i$ be the induced matching of $M$ on $A_i$. Note that $M_i$ is also the induced matching of $M'$ on $A_i$.
	For each index $i$ we add edges $B_i$ on $A_i$ which do not cross any edge in $M_i$ such that $M_i$ and $B_i$ do not have an edge in common and $B_1 \cup M_i$ is a triangulation of $A_i$.
	
	We claim that $B_i$ spans all points in $A_i$.
	Clearly, $B_i \cup M_i$ spans all points in $A_i$.
	Let $e$ be an edge of $M_i$ and $\Delta$ be a triangle of $B_1 \cup M_i$ that contains $e$.
	Since $M_i$ is a matching, $\Delta$ and $M_i$ have exactly $e$ in common.
	Hence, $B_i$ contains the other two edges of $\Delta$.
	Removing any edge $e$ from $B_i \cup M_i$ does not lead to a disconnected drawing.
	Therefore, $B_i$ spans all points in $A_i$.
	
	Merging all the drawings $B_i$ and $B'_j$ we get a spanning drawing on $S$.
	Breaking cycles one by one we eventually obtain a spanning tree.
\end{proof}

\begin{figure}[htb]
	\vspace{-3ex}
	\centering
	\includegraphics[page=1,scale=1]{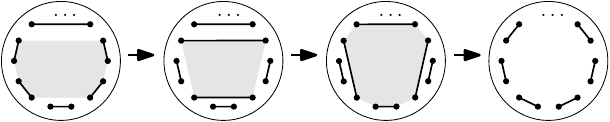}
	\caption{Rotation of a 6-ear in 3 steps (in each step we rotate the grey inside cycle).}
	\label{fig:long-cycles-maintext}
	\vspace{-3ex}
\end{figure}

\smallskip
We next consider sufficiently large ears. The following lemma states that such ears can be rotated in at most three steps; see \cref{fig:long-cycles-maintext} for a sketch of this sequence of rotations, whose proof uses \cref{insidecycle}.
Note that \cref{large_cycles} also implies that the two perimeter matchings have distance at most 3 in $\dcg(\tree)$.

\begin{restatable}{lemma}{lemmaLargeCycles}\label{large_cycles}
	 Let $M$ and $M'$ be two matchings whose symmetric difference is a $k$-ear with $k\ge 6$. Then $M$ and $M'$
	have distance at most 3 in $\dcg(\tree)$.
\end{restatable}

\begin{proof} 
	The idea of the proof is to perform three rotations of inside cycles. Each rotation can be done in one step due to Lemma~\ref{insidecycle}.
	We proceed as in \cref{fig:long-cycles}: 
	\begin{figure}[htb]
		\centering
		\includegraphics[page=1,scale=1]{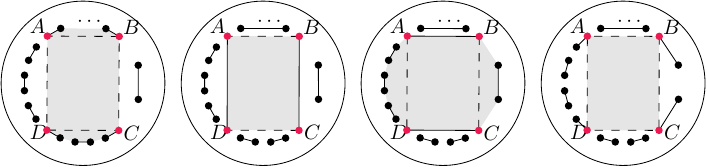}
		\caption{Intermediate steps for the rotation of a k-ear with $k\geq6$.}
		\label{fig:long-cycles}
	\end{figure} \\
	First we find 4 points $A$, $B$, $C$, $D$ of the ear such that each of the four arcs $\widehat{AB}$, $\widehat{BC}$, $\widehat{CD}$, $\widehat{DA}$ of the ear contain a positive even number of points in its interior. Without loss of generality the points $A$, $B$ are matched inside $\widehat{AB}$ and $C$, $D$ are matched inside $\widehat{CD}$. We do the following three steps: Rotate $\widehat{AB}\widehat{CD}$, rotate a 2-cycle $ABCD$, rotate $\widehat{BC}\widehat{DA}$. Since each arc initially contained at least two points, each step rotates an inside cycle and it is easily checked that this transforms $M$ into $M'$.
\end{proof}

\begin{restatable}{theorem}{thmUpperBound}\label{upper_bound}
	For $2n\ge 10$, the graph $\dcg(\tree)$ is connected with diameter $\diam(\dcg(\tree))\le 5$.
\end{restatable}

\begin{proof}   
	For $2n=10$, the statement follows from checking all pairs of matchings.
	\Cref{fig:g10-text} gives a schematic depiction of the whole graph $\dcg(\tree)$ in this case. 
	If we want to find a path between rotated version of some nodes,
	we just need to find a walk in the picture along which 
	the rotations compose to the desired value.
	
	\begin{figure}[htb]
		\centering
		\includegraphics[page=1,scale=1]{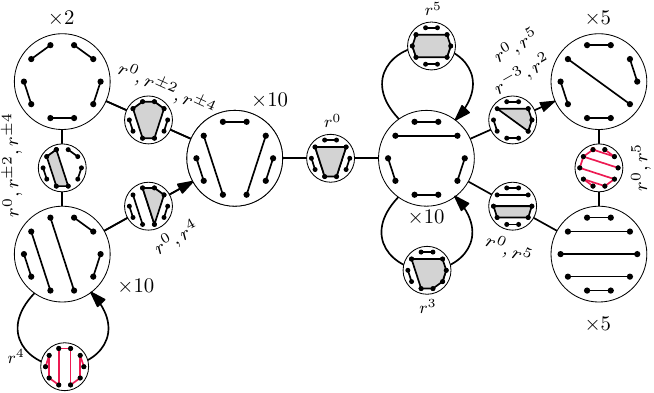}
		\caption{Schematic depiction of the whole graph $G_{10}$. The letter $r$ stands for a possible rotation by $2\pi/10$. Going against the arrows rotates in opposite direction.
			Next to each vertex, the number of different matchings resulting from rotations is indicated.
			The edges indicate either the rotation of an inside cycle (in gray),
			or a compatible spanning tree (red).}
		\label{fig:g10-text}
	\end{figure}
	
	Now assume $2n\ge 12$.
	We color the perimeter alternately in blue and red and refer to the odd (resp.\ even) perimeter matching as the blue perimeter matching $B$ (resp.\ red perimeter matching $R$). Moreover, for a fixed matching $M$, let $\dm(M)=\min\{\d(M,B),\d(M,R)\}$ and $\dM(M)=\max\{\d(M,B),\d(M,R)\}$ be the distance from $M$ to the closer and the further perimeter matching, respectively.
	Since by \cref{large_cycles} we have \mbox{$\d(B,R)\le 3$}, it suffices to show that the non-perimeter matchings can be split into three classes $A_1$, $A_2$, $A_3$ with the following properties (see \cref{fig:distances}):
	\begin{enumerate}
		\item[1.] $\forall M\in A_1$ we have $\dm(M)\le 1$ (and hence $\dM(M)\le 1+3=4$); 
		\item[2.] $\forall M\in A_2$ we have $\dm(M)\le 2$ and $\dM(M)\le 3$;
		\item[3.] $\forall M\in A_3$ we have $\dM(M)\le 3$ and $\forall M,M'\in A_3$ we have \mbox{$\d(M,M')\le 4$}.
	\end{enumerate}
	
	\begin{figure}[htb] 
		\centering
		\includegraphics[page=1,scale=1]{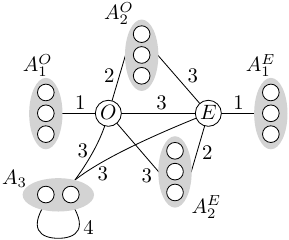}
		\caption{Depiction of the partitioning of the set of all non-perimeter matchings into subsets $A_1=A_1^E \cup A_1^O$, $A_2=A_2^E \cup A_2^O$, and $A_3$, with bounds on their distances. For $i\in \{1,2\}$, the matchings in $A^E_i$ and $A^O_i$ have a smaller upper bound on their distance to $E$ and $O$, respectively.}
		\label{fig:distances}
	\end{figure}
	
	Fix a matching $M$. It consists of a number (possibly zero) of diagonals, odd perimeter edges (shown in blue), and even perimeter edges (shown in red).
	The convex hull of $S$ is split by the diagonals into several polygons, each of them corresponding to a cycle. The dual graph $D(M)$ of these polygons is a tree. Its leaves correspond to ears and the interior nodes correspond to inside cycles.
	Since the diagonals of $M$ split the perimeter into (possibly empty) arcs that alternately consist of only red and only blue sides, the nodes of the tree can be properly two-colored in blue and red by the color of the perimeter edges of the corresponding polygons (see \cref{fig:quantb}).
	
	\begin{figure}[htb] 
		\centering
		\includegraphics[page=1,scale=1]{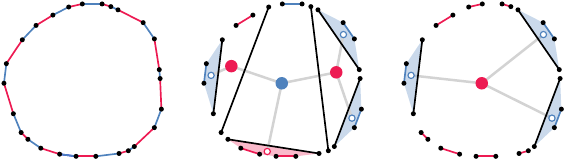}
		\caption{For a fixed matching $M$, we color the perimeter edges alternately in blue (odd edges) and red (even edges). The coloring extends to a proper coloring of a tree $D(M)$ that is dual to $M$. In the shown example, rotating the inside cycle corresponding to the blue interior node of $D(M)$ creates a tree which leaves all have the same color.
		}
		\label{fig:quantb}
	\end{figure}
	
	Now we distinguish four cases based on what the dual tree $D(M)$ looks like. Let $b$ and $r$ be the number of leaves in $D(M)$ colored blue and red, respectively. Without loss of generality we assume that $b\ge r$. Remember that by Lemma~\ref{insidecycle} we can rotate any number of disjoint inside cycles in one step.
	\begin{itemize}
		\item [$\bullet$]\underline{$b\ge 1,r=0$}: If $b=1$ then $M=B$. Otherwise, we simultaneously rotate all red inside cycles. This removes all diagonals, we reach $B$ in 1 step and we put $M$ into $A_1$.
		\item [$\bullet$]\underline{$b\ge 2, r\ge 2$}: We can get to $B$ in 2 steps: First, simultaneously rotate all blue inside cycles (this removes all diagonals except the ones separating blue leaves of $D(M)$). Then rotate the (only, red) inside cycle. Similarly, we can reach $R$ in 2 steps, hence $M$ can go to $A_2$. (This case can only occur when $2n\ge 16$.)
		\item [$\bullet$]\underline{$b\ge 2, r=1$}: See \cref{fig:quant-3nodes}. In the first step, rotate all blue inside cycles to get $b\ge 2$ blue leaves and one (red) inside cycle. To get to $B$, rotate the inside cycle ($\le 2$ steps total). To get to $R$, note that the original diagonal that cut off the red leaf disappeared in the first step, hence it was rotated out and we must now have at least $1+1+1\ge 3$ consecutive red sides, say $e$, $f$, $g$.
		\begin{figure}[htb]
			\centering
			\hspace{0.5cm}\includegraphics[page=1,scale=1]{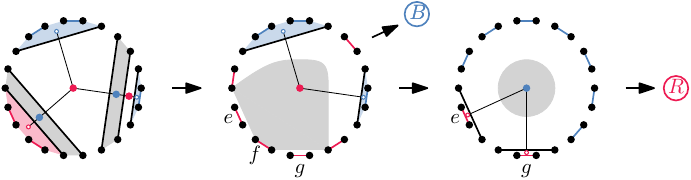}
			\caption{When $b\ge 2$ and $r=1$ we can get to $B$ in 2 steps and to $R$ in 3 steps.}
			\label{fig:quant-3nodes}
		\end{figure}
		Rotate the inside without $e$ and $g$ and then rotate the inside. This gets to $R$ in 3 steps, hence $M$ can go to $A_2$. (This case can only occur when $2n\ge 14$.)
		
		\item [$\bullet$]\underline{$b=1,r=1$}: In the first step, rotate all blue inside cycles and push the diagonal that cuts off the blue leaf to a side, if it is not there yet, by rotating the whole blue ear without one blue perimeter edge (see \cref{fig:quant-2nodes}(a)). Since $2n\ge 10$, we have at least 3 consecutive red edges and, as in the previous case, we can thus reach $R$ in two more steps (for a total of 3 steps). Likewise for $B$, hence we aim to put $M$ into $A_3$.
		\begin{figure}[htb]
			\centering
			\hspace{0.5cm}\includegraphics[page=1,scale=1]{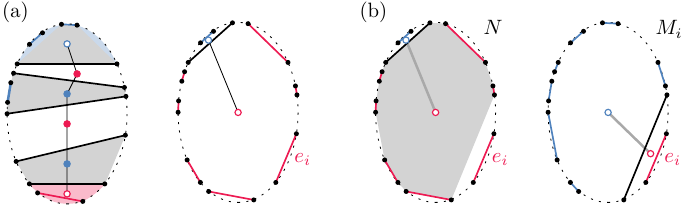}
			\caption{Intermediate steps for the case $b=1$ and $r=1$.}
			\label{fig:quant-2nodes}
		\end{figure}
		
		For that, we need to check that any two such matchings are distance at most 4 apart. To that end, it suffices to check that any two matchings $N$, $N'$ with one diagonal that cuts off a single blue perimeter edge are in distance at most $4-1-1=2$ apart. This is easy (see \cref{fig:quant-2nodes}(b)): Label the $n$ red perimeter edges by $e_1,\dots,e_n$ and for each $i=1,\dots,n$, denote by $M_i$ the matching with one diagonal that cuts off the perimeter edge $e_i$. We claim that  some $M_i$ is adjacent to both $N$ and $N'$. In fact, we claim that $N$ is adjacent to at least $n-2$ of the $n$ matchings $M_i$. Indeed, for any of the $n-2$ red sides $e_i$ present in $N$, we can rotate the (inside) cycle consisting of the red leaf of $D(N)$ without $e_i$. The same holds for $N'$. Since for $2n\ge 10$, we have $(n-2)+(n-2)>n$, there is a matching $M_i$ adjacent to both $N$ and $N'$.
	\end{itemize}
\end{proof}

\subsection{A lower bound for the diameter of $\dcg(\tree)$}

\begin{figure}[htb]
	\centering
	\includegraphics[page=1,scale=1]{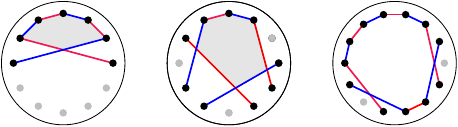}
	\caption{Boundary areas with five points (left) and three points (middle). The drawing on the right does not show a boundary area; not all points are neighboring on the convex hull of $S$.}
	\label{fig:boundaryarea}
\end{figure}

Since the diameter of $\dcg(\tree)$ has a constant upper bound, is seems reasonable to also ask for a best possible lower bound.
To do so, we first identify structures which prevent that two matchings are $\tree$-compatible.
	Let $M$ and $M'$ be two matchings in $S$. A \emph{boundary area with $k$ points} is an area within the convex hull of $S$ containing $k$ points of $S$ that is bounded by edges in $M$ and $M'$ such that these edges form at least one crossing and such that all points of $S$ on the boundary of the area form a sequence of consecutive points of $S$ along the boundary of the convex hull of $S$; see \cref{fig:boundaryarea}.

\begin{figure}[htb]
	\centering
	\includegraphics[page=1,scale=1]{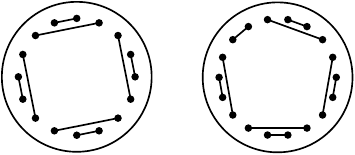}
	\caption{Left: A 2-semiear matching. Right: A near-2-semiear matching.}
	\label{fig:twoearneartwoear}
\end{figure}

We next define two special matchings.
	A \emph{2-semiear matching} is a matching on a set of $4k$ points consisting of exactly $k$ 2-semiears and an inside $k$-semicycle (with all its edges being diagonals).
	Similarly, a \emph{near-2-semiear matching} is a matching on a set of $4k+2$ points consisting of exactly $k$ 2-semiears and an inside $(k+1)$-semicycle;
	see  \cref{fig:twoearneartwoear}.

As for perimeter matchings,  we distinguish between \emph{odd} and \emph{even} \mbox{(near-)} 2-semiear matchings.
If the perimeter edges of the 2-semiears are labeled 'even' then we call the (near-)2-semiear matching even, otherwise we call it odd.

\begin{restatable}{lemma}{lemmaEarAndBoundaryArea}\label{earandboundaryarea}
	Let $M$, $M'$ be two matchings whose symmetric difference is an ear or a boundary area with at least three points. Then $M$ and $M'$ are not $\tree$-compatible to each other.
\end{restatable}

\begin{figure}[htb]
	\centering
	\includegraphics[page=1,scale=1]{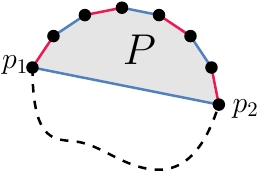}
	\caption{Two matchings $M$ and $M'$ (depicted in red and blue) creating an ear. The points $p_{1}$ and $p_{2}$ might be connected by a spanning tree outside the ear.}
	\label{fig:alternatingear}
\end{figure}

\begin{proof}
	We consider two matchings $M$ and $M'$ creating a $k$-ear and we call the respective polygon $P$ (cf.\ \cref{fig:alternatingear}). The proof for a boundary area with at least three points works in a similar way.  
	
	If the two matchings are $\tree$-compatible, we can draw an edge-disjoint tree in $S$.
	Let $p_{1}$ and $p_{2}$ be the two endpoints of the diagonal in the ear. Any other point in $P$ cannot be directly connected to a point outside $P$ via a tree edge, therefore at least $k-2$ tree edges need to lie within $P$ (if $p_{1}$ and $p_{2}$ are connected to each other outside $P$; otherwise even $k-1$ tree edges are needed). 
	However, by planarity there can be at most $k-3$ edges in a polygon spanned by $k$ points, a contradiction.
\end{proof}

\begin{restatable}{lemma}{lemmaEarMatchingCompatible}\label{2earmatchingcompatible}
	Let $M$ be a matching that is $\tree$-compatible to an even (odd) 2-semiear-matching. Then $M$ contains no odd (even) perimeter edge.
\end{restatable}

\begin{figure}[htb]
	\centering
	\includegraphics[page=1,scale=1]{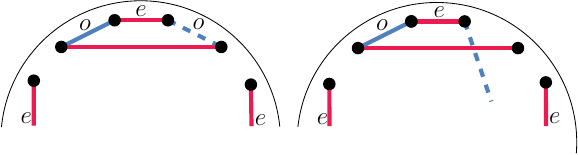}
	\caption{An (even) 2-semiear matching drawn in red and a blue matching with at least one odd perimeter edge; on the left the blue matching creates a cycle with the red matching, on the right a boundary area with three points occurs.}
	\label{fig:twocycles}
\end{figure}

\begin{proof}
	We prove the statement by contradiction and assume that there exists a $\tree$-compatible matching $M$ which contains at least one odd perimeter edge.
	This matching edge connects one endpoint of a perimeter edge with its neighboring vertex (matched by a diagonal in the 2-semiear matching), see \cref{fig:twocycles}. 
	We distinguish between the cases where the other endpoint of the perimeter edge is matched to (in $M$). If it is matched with the same diagonal of the 2-semiear, the two matchings create an ear, a contradiction to \cref{earandboundaryarea} (cf.\ \cref{fig:twocycles} on the left). 
	Otherwise, this matching edge intersects with the diagonal of the 2-semiear matching, thus creates a boundary area with three points, a contradiction to \cref{earandboundaryarea} (cf.\ \cref{fig:twocycles} on the right).	
\end{proof}

The following lemma can be proven in a similar way.

\begin{restatable}{lemma}{lemmaNearEarMatchingCompatible}\label{near2earmatchingcompatible}
	Let $M$ be a matching that is $\tree$-compatible to a near-2-semiear-matching $M'$ consisting of $k$ even (odd) and one odd (even) perimeter edge. Then $M$ contains at most one odd (even) perimeter edge, which is the one in $M'$.
\end{restatable}

\begin{proof}
	Let $M'$ be a matching $\tree$-compatible to a near-2-semiear matching $M$ as defined in the statement. 
	All but one of the odd perimeter edges would connect an even perimeter edge with its diagonal in an 2-semiear, therefore cannot be contained in $M'$ as shown in the proof of \cref{2earmatchingcompatible}.
	Consequently, there is at most one odd perimeter edge in $M'$ (which is exactly the odd perimeter edge in $M$).
\end{proof}

\begin{restatable}{lemma}{lemmaEars}\label{ears}
	Let $M$ and $M'$ be two $\tree$-compatible matchings. Then $M$ and $M'$ have at least two perimeter edges in common.
\end{restatable}

\begin{proof} 
	Let $M'$ be a matching $\tree$-compatible to $M$.
	First of all, we consider the case that $M$ is a perimeter matching. Without loss of generality, $M$ is the even perimeter matching.
	Our claim is that $M'$ has no odd semiear.
	Assume to the contrary that $M'$ has odd semiears.
	Any odd semiear in $M'$ creates an ear with $M$, thus the two matchings cannot be $\tree$-compatible to each other, a contradiction to the assumption.
	Therefore we can conclude that our statement holds for perimeter matchings since every non-perimeter matching contains at least two semiears. (Consider the dual graph where the areas defined by matching edges correspond to points and two points are connected if and only if the two areas are separated by a matching edge. This graph forms a tree where semiears in the matching correspond to leaves in the tree.)
	
	All other matchings have at least two semiears and we distinguish different cases.
	\begin{itemize}
		\item []\underline{Case 1: There exist two semiears of size $\geq3$ in $M$} \\
		Our claim is that at least one of the perimeter edges of each semiear lies in $M'$. We consider one semiear and  assume to the contrary that none of the perimeter edges of this semiear lies \mbox{in $M'$}. \\
		Without loss of generality we assume that the semiear is even.
		Thus by assumption, every vertex of this semiear is either matched by an odd perimeter edge or by a diagonal in $M'$.
		If all points in the semiear are matched by odd perimeter edges in $M'$, we get an ear contradicting \cref{earandboundaryarea} (cf.\ \cref{fig_app:earscase1} (a)). \\
		If two points in the semiear are matched with each other by a diagonal (in $M'$), the other points (in the semiear) are separated into two sets. Those on the side with just perimeter edges have to be matched with each other in $M'$, otherwise $M'$ would intersect itself. We can iteratively shrink this side, until the remaining points are all matched by odd perimeter edges. This again creates an ear (cf.\ \cref{fig_app:earscase1} (b)). \\
		Otherwise, at least one diagonal in $M'$ intersects the diagonal (in $M$) of the semiear, starting at an endpoint of an even perimeter edge. If the other endpoint of this edge is matched by an odd perimeter edge in $M'$, we get a boundary area with at least four points, therefore no spanning tree can be drawn and the matchings are not $\tree$-compatible (cf.\ \mbox{\cref{fig_app:earscase1} (c)).} \\
		If the other endpoint of the even perimeter edge is also matched by a diagonal in $M'$, we get a so called 'blocking structure', i.e.,\ the two endpoints of the perimeter edge cannot be connected directly by a spanning tree.
		Since we already excluded diagonals within the semiear, the vertex neighboring this perimeter edge also has to be matched by a diagonal in $M'$ (and it exists since we assumed that the size of the semiear is at least 3). We again consider the other endpoint of this even perimeter edge and either construct a boundary area with at least three points (cf.\ \cref{fig_app:earscase1} (d)), which again leads to a contradiction or we get a second blocking structure (cf.\ \cref{fig_app:earscase1} (e)). However, this concludes this case as well, since the points inbetween the two blocking structures are separated from the other points and cannot be connected with them by any spanning tree.
		
		It follows that at least one of the perimeter edges in the semiear of $M$ also lies in $M'$. 
		Analogously we can apply the argument for the other semiear.\\		
		
		\begin{figure}[tb]
			\centering
			\includegraphics[page=1,scale=1]{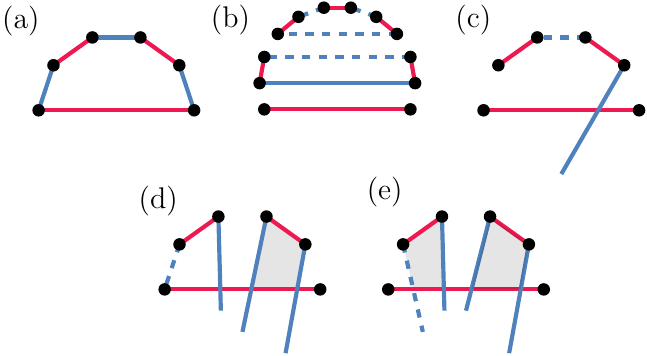}
			\caption{All possible cases for a semiear of size $k\geq 3$ in a matching $M$ (depicted in red) and a second matching $M'$ (depicted in blue) which does not use any of the perimeter edges in $M$}
			\label{fig_app:earscase1}
		\end{figure}

		\item[]\underline{Case 2: All but one semiear in $M$ is of size 2} \\ 
		For simplicity we assume without loss of generality that there exists an even 2-semiear in $M$. Matching the points of this semiear by odd perimeter edges yields an ear, a contradiction by \cref{earandboundaryarea} (cf.\ \cref{fig_app:earscase2} (a)). 
		If one of the endpoints of the even perimeter edge is matched by an odd perimeter edge in $M'$ and the other one is matched by a diagonal, we get a boundary area with three points, contradicting \cref{earandboundaryarea} (cf.\ \cref{fig_app:earscase2} (b)). \\
		Therefore, both endpoints of the even perimeter edge are matched by diagonals of $M'$ which intersect the diagonal of the 2-semiear (cf.\ \cref{fig_app:earscase2} (c)). 
		We can assume that this holds for all semiears of size two in $M$, otherwise we apply one of the arguments above. \\
		
		\begin{figure}[htb]
			\centering	
			\includegraphics[page=1,scale=1.3]{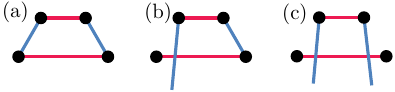}
			\caption{All possible cases for a 2-semiear in a matching $M$ (depicted in red) and a second matching $M'$ (depicted in blue) which does not use the perimeter edges in $M$}
			\label{fig_app:earscase2}
		\end{figure}
		
		Out of the 2-semiears of $M$ we choose the one with no further semiear of $M$ (also not the one of larger size) on one side of a diagonal $d$ in $M'$, where $d$ is incident to a point of the perimeter edge of that ear. This is possible since the number of semiears is finite and the diagonals in $M'$ cannot intersect each other, therefore there is an ordering of the 2-semiears in $M$ (and only one semiear of larger size). 
		Without loss of generality there is no semiear of $M$ left of $d$.
		It is easy to see that the diagonal $d$ induces a semiear in $M'$ left of $d$. If this semiear is of size 2 and two diagonals $d'_1$ and $d'_2$ in $M$ are intersecting the diagonal of the semiear, we get another blocking structure. 
		(Otherwise we can apply one of the other arguments above to the semiear in $M'$ and again end up with a perimeter edge lying in both matchings.) It follows that $d'_1$ and $d'_2$ have to intersect the diagonals in $M'$ that intersect the even 2-semiear. Otherwise another semiear in $M$ to the left of $d$ would be induced, a contradiction. However, this separates at least three points from the rest and it is not possible to find a common compatible spanning tree (cf. \cref{fig:2earblocking}). \\
		
		\begin{figure}[tb]
			\centering
			\includegraphics[page=1,scale=1]{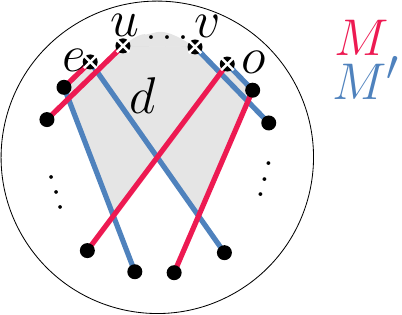}
			\caption{An even 2-semiear in $M$ (red matching edges) intersected by two diagonals in $M'$ (blue edges) (on the left) and an odd 2-semiear in $M'$ intersected by two diagonals in $M$ (on the right). The vertices $u$ and $v$ might coincide. The grey areas are blocked, i.e.,\ the spanning tree cannot pass them, therefore at least three points (if $u=v$) are not reachable from the rest of the vertices (marked by white crosses). }
			\label{fig:2earblocking}
		\end{figure}
		
		\item[]\underline{Case 3: All semiears in $M$ are of size 2} \\
		This case works similar to the second case. If the cases (a) or (b) in \cref{fig_app:earscase2} can be applied to two 2-semiears, we are done since both perimeter edges also lie in $M'$. If we can apply one of those cases to at least one 2-semiear, we treat this semiear like the semiear of larger size in Case 2 and proceed as before.\\
		Otherwise, all 2-semiears in $M$ are as depicted in \cref{fig_app:earscase2} (c). Again there is an ordering of those 2-semiears and now we can choose two of them such that there is no further semiear of $M$ on one side of a diagonal in $M'$. (Once there is no further semiear on the 'left' side, once there is no semiear on the 'right' side.) It follows that two distinct semiears in $M'$ are induced. The arguments in Case 2 can be applied separately to both of them, therefore we end up with at least two perimeter edges which lie in both $M$ and $M'$.
	\end{itemize}
\end{proof}

\begin{restatable}{corollary}{corLowerBound}\label{lowerbound}
	Let $S$ be of size $2n\geq 10$.
	For even $n$, the distance between an even 2-semiear matching and an odd 2-semiear matching is at least 4.\\
	For odd $n$, let $M$ be a near-2-semiear matching with a single even perimeter edge $e$ and let $M'$ be a near-2-semiear matching with a single even perimeter edge $e'$
	that shares a vertex with $e$.
	Then the distance between $M$ and $M'$ is at least 4.
\end{restatable}

\begin{figure}[htb]
	\centering
	\includegraphics[page=1,scale=1]{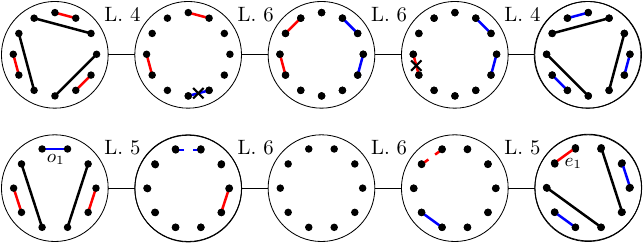}
	\caption{Illustrations, that the distance between two special 2-semiear matchings (left) and between two special near-2-semiear matchings (right) is at least 4. Even perimeter edges are drawn in red, odd ones are drawn in blue. The numbers next to the edges indicate which Lemma is applied. Crossed out edges indicate that this type of edge (even or odd) cannot appear in that matching.}
	\label{fig:app:diameter_4k_4k+2}
\end{figure}

\begin{proof}
	\underline{$n$ is even:} \\
	By \cref{2earmatchingcompatible} we know that for every matching $\tree$-compatible to an even 2-semiear matching all perimeter edges are even. Now by \cref{ears} all matchings which are $\tree$-compatible to them contain at least two of their even perimeter edges. \\
	Analogoulsy, in every matching $\tree$-compatible to an odd 2-semiear matching all perimeter edges are odd, and all matchings $\tree$-compatible to those contain at least two odd perimeter edges (in particular any matching with no odd perimeter edge is not $\tree$-compatible). \\
	Combining this results shows that there are at least three intermediate matchings between an even and an odd 2-semiear matching in the disjoint $\tree$-compatible graph.
	
	\underline{$n$ is odd:} \\
	By \cref{near2earmatchingcompatible} every matching $\tree$-compatible to $M$ contains at most one odd perimeter edge, namely the same as in $M$, say $o_{1}$. Analogously, every matching $\tree$-compatible to $M'$ contains no  even perimeter edge other than the one in $M'$, say $e_{1}$. \\
	As before we can apply \cref{ears} and deduce that all matchings $\tree$-compatible to those with at most one odd or even perimeter edge, respectively, contain at least two perimeter edges. However, since $o_{1}$ and $e_{1}$ are incident, they cannot both appear in any of the $\tree$-compatible matchings at the same time, thus the two sets of all $\tree$-compatible matchings is disjoint which implies a total lower bound of four for the distance of $M$ and $M'$.
\end{proof}

\section{Disjoint caterpillar-compatible matchings}\label{sec:caterpillar}
A natural question is what happens if we do not take the set of all plane spanning trees, but a smaller set.

A \emph{caterpillar}
(from $p$ to $q$) is a tree which consists of a path (from $p$ to $q$, also called \emph{spine}) and edges with one endpoint on the path.
These latter edges
are also called the \emph{legs} of the caterpillar.
We denote the set of all plane spanning caterpillars by $\cater$.
Furthermore, a \emph{one-legged caterpillar} is a caterpillar where every vertex of the spine is incident to at most one leg.
We denote the family of all plane spanning one-legged caterpillars in $S$ by $\cater_3$.
Note that every vertex of a one-legged caterpillar has degree at most 3.
Hence, one-legged caterpillars are special instances of trees with maximum degree 3.

\begin{restatable}{lemma}{lemmaCompCaterpillar}\label{lemma:caterpillar}
	For any edge $e=pq$ of a matching $M$ there exists a plane one-legged caterpillar compatible to $M$ from $p$ to $q$ which spans all points between $p$ and $q$ along the boundary of the convex hull of $S$ (on either side of $e$). This caterpillar has vertices of at most degree $3$ and $p$ and $q$ are vertices of its spine.
\end{restatable}

\begin{figure}[htb]
	\centering
	\vspace{-3ex}
	\includegraphics[page=1,scale=1]{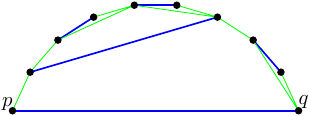}
	\caption{A matching (in blue) and a compatible caterpillar (in green) constructed in the proof of \cref{lemma:caterpillar}.}
	\label{fig:caterpillar_edge}
\end{figure}

\begin{proof} 
	We construct the caterpillar $C$ in a greedy way from $p$ to $q$.
	At the start, let $C$ be the point $p$.
	Assume we have a one-legged caterpillar $C$ from $p$ to a point $x$, and $C'$ contains each point between $p$ and $x$. 
	Let $y$ and $z$ be the next two points from $x$ to $q$.
	If $xy$ is not an edge of $M$, we add $xy$ to $C$ and continue from $y$.
	Otherwise, if $xy$ is an edge of $M$, then $xz$ and $yz$ are not edges of $M$.
	We add $xz$ and $yz$ to $C$ and continue from~$z$.
	By construction, every spine vertex has at most one leg.
	Further, every point between $p$ and $q$ is in $C$ by construction.
	So we constructed a one-legged caterpillar.
	An example is depicted in \cref{fig:caterpillar_edge}.
\end{proof}

Note that every matching $M$ contains a perimeter edge and by \cref{lemma:caterpillar} there also exists a caterpillar which is disjoint compatible to $M$.
Further, by construction $p$ is incident to only one edge.

\begin{restatable}{lemma}{lemmaCatInside}\label{lemma:cat_inside}
	Let $M$ and $M'$ be two matchings whose symmetric difference is an inside cycle. Then $M$ and $M'$ are disjoint $\cater$-compatible.
\end{restatable}

\begin{proof}
	Let $K$ be the inside cycle which is the symmetric difference of two perfect matchings $M$ and $M'$.
	For every diagonal of $K$ we construct a caterpillar as in \cref{lemma:caterpillar} and merge caterpillars which have a point in common.
	If every edge of $K$ is a diagonal, then merging all those caterpillars gives a cycle consisting of the spine edges. By deleting one spine edge we obtain a spanning caterpillar.
	Otherwise, the merging yields a set $C_1, \dots, C_r$ of caterpillars whose endpoints are points of the diagonals.
	We label the two  endpoint of $C_i$ with $s_i$ and $t_i$ in clockwise direction; cf~\cref{fig:disj_cater_gen}.
	Note that $s_it_i$ might not be a diagonal.
	Further note that every point of $S$, which is not on $K$, is in one of $C_1, \dots, C_r$.
	
	\begin{figure}[htb]
		\centering
		\includegraphics[page=1,scale=1]{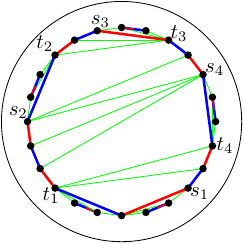}
		\caption{Constructed caterpillar (in green) for two disjoint $\cater$-compatible matchings (in red and blue, common edges in blue).
			Note that one red diagonal and one blue diagonal have a point in common.}
		\label{fig:disj_cater_gen}
	\end{figure}
	
	We connect these caterpillars in the following way.
	We add the edges $t_is_{r+1-i}$ for $1 \leq i\leq \frac{r}{2}$, and $s_{i-1}t_{r+1-i}$ for $2 \leq i\leq \frac{r}{2}$.
	This gives us a caterpillar $C$ which contains all $C_i$.
	Note that $C$ starts at $s_1$ and ends at $s_{r/2}$ if $r$ is even, and at $t_{(r+1)/2}$ if $r$ is odd.
	
	For any point $p$ between $t_r$ and $s_1$ we add the edge $t_1p$ to $C$.
	If $r$ is odd, then we add the edge $s_{r+1-i}p$ to $C$ for any point $p$ between $t_i$ and $s_{i+1}$, where $1\leq i \leq r-1$.
	$C$ ends at point $t_{(r+1)/2}$ since $r$ is odd. Hence, all these edges are on the inside of $K$.
	
	If $r$ is even, then we add the edge $s_{r+1-i}p$ to $C$ for any point $p$ between $t_i$ and $s_{i+1}$, where $1\leq i \leq r-1$ and $i\neq \frac{r}{2}$.
	For any point $p$ between $t_{r/2}$ and $s_{r/2 +1}$, we add the edge $s_{r/2 +1}p$ to $C$.
	
	After adding all these edges to $C$, any point of $K$ is contained in $C$.
	Hence, $C$ is a spanning caterpillar.
\end{proof}

Note that \cref{lemma:cat_inside} is a sufficient condition for $\cater$-compatibility of matchings similar to \cref{insidecycle} for $\tree$-compatibility.
Adapting the proof of \cref{upper_bound} to rotate only one cycle (instead of several) per step, and noting that the number of cycles is $O(n)$, we get the following theorem.

\begin{restatable}{theorem}{lemmaCaterpillarConnected}\label{caterpillar-compatible}
	For $2n\geq 10$, the graph $\dcg(\cater)$ is connected with diameter $\diam(\dcg(\cater))=\O(n)$.
\end{restatable}

\begin{proof}
	For $2n=10$, \cref{fig:g10-text} gives a schematic depiction of $\dcg(\tree)$. Note that either only one inside cycle is rotated or the indicted tree is a caterpillar.
	Hence, \cref{lemma:cat_inside} is also a schematic depiction of $\dcg(\cater)$ for $2n=10$.
	
	In the proof of \cref{upper_bound} we showed for $2n\geq 12$ that $\diam(\dcg(\tree))\leq 5$ by rotating multiple inside cycles at once.
	Hence, for any two matchings $M$ and $M'$ there exist a sequence of matchings $M=M_0, M_1, M_2, M_3, M_4, M_5=M'$ such that the symmetric difference of$M_i$ and $M_{i+1}$ is a set of inside cycles for $0\leq i \leq 4$.
	Now consider two matchings $M_i$ and $M_{i+1}$ whose symmetric difference is a set of inside cycles.
	Note that the number of inside cycles is at most $n/2$ since each of these cycles contains at least $4$ points.
	By \cref{lemma:cat_inside} we can rotate one inside cycle in one step.
	Therefore, $M_{i}$ and $M_{i+1}$ are at distance at most $n/2$ in $\dcg(\cater)$ for any $0\leq i \leq 4$.
	It follows that $\diam(\dcg(\cater))\leq 5n/2$.
\end{proof}

Next we consider disjoint $\cater_3$-compatible matchings.
As before, we first find a sufficient condition for their compatibility.

\begin{restatable}{lemma}{lemmaOneLeggedCatInside}\label{lemma:oneleggedcat_inside}
Let $M$ and $M'$ be two matchings whose symmetric difference is an inside $2$-cycle. Then $M$ and $M'$ are disjoint $\cater_3$-compatible.
\end{restatable}

\begin{figure}[htb]
	\centering
	\includegraphics[page=1,scale=1]{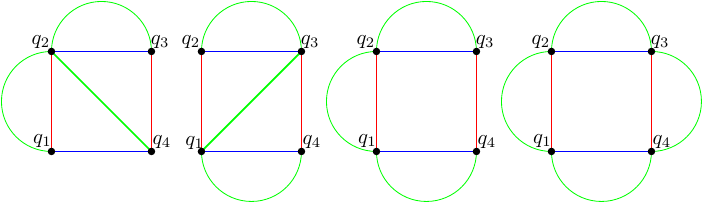}
	\caption{All possibilities for an inside $2$-cycle (drawn as red and blue square) with a disjoint compatible caterpillar sketched (in green). The half circles are caterpillars.}
\label{fig:caterpillar_2-cycles}
\end{figure}

\begin{proof}
We have four cases depending on how many edges of the $2$-cycle are diagonals and their relative position.
The four cases are depicted in \cref{fig:caterpillar_2-cycles}.

In the leftmost case, where
exactly two diagonals share a point, we take two one-legged caterpillars from $q_2$ to $q_1$ and $q_3$, respectively, constructed as in the proof of \cref{lemma:caterpillar}.
Note that each such caterpillar has degree $1$ at its start point. Hence, together with the edge $q_2q_4$ they form a one-legged caterpillar which is disjoint compatible to both $M$ and $M'$.

If we have two diagonals which are not adjacent, we take two one-legged caterpillars from $q_2$ to $q_3$ and from $q_1$ to $q_4$, constructed as in the proof of \cref{lemma:caterpillar}.
We connect these two caterpillars with the edge $q_1q_3$ and obtain a spanning one-legged caterpillar.

If we have three diagonals, we take the one-legged caterpillars from $q_1$ to $q_2$, from $q_2$ to $q_3$  and from $q_3$ to $q_4$, constructed as in the proof of \cref{lemma:caterpillar}.
This is already a spanning one-legged caterpillar.

If we have four diagonals, we take the one-legged caterpillars from $q_1$ to $q_2$, from $q_2$ to $q_3$, from $q_3$ to $q_4$ and from $q_4$ to $q_1$, constructed as in the proof of \cref{lemma:caterpillar}.
The spines of these caterpillars form a cycle. Deleting any of the spine edges yields a spanning one-legged caterpillar.
\end{proof}

With this, we can show the following theorem.

\begin{restatable}{theorem}{lemmaOneLegCaterpillarConnected}
	For $2n\geq 10$, the graph $\dcg(\cater_3)$ is connected and has diameter $\diam(\dcg(\cater_3))=\O(n)$.
\end{restatable}

\begin{proof}
	We first show that any two matchings $M$ and $M'$ whose symmetric difference is a single inside cycle $K$ are connected in $\dcg(\cater_3)$.
	
	Consider an inside cycle $K$ and label its points with $u_0, u_1, \dots u_x, v_y, v_{y-1}, \dots v_0$ such that $u_0v_0$ and $u_xv_y$ are two diagonals of $K$.
	Note that since $K$ has an even number of edges the parity of $x$ and $y$ is the same.
	We split $K$ into interior-disjoint $2$-cycles $K_i$ in the following way; cf~\cref{fig:2-cycle-decomposition_app}.
	
	\begin{figure}[htb]
		\centering
		\includegraphics[page=1,scale=1]{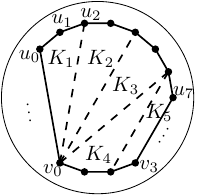}
		\caption{Subdivision of an inside $6$-cycle into five inside $2$-cycles.}
		\label{fig:2-cycle-decomposition_app}
	\end{figure}
	If $x$ is even, then $K_i$ is the $2$-cycle $v_0u_{2i}u_{2i+1}u_{2i+2}$ for $0\leq i \leq \frac{x}{2}-1$. 
	For $\frac{x}{2} \leq i \leq \frac{x+y}{2}-1$, let $K_i$ be the $2$-cycle $u_xv_{2i-x}v_{2i-x+1}v_{2i-x+2}$.
	
	If $x$ is odd, then $K_i$ is the $2$-cycle $v_0u_{2i}u_{2i+1}u_{2i+2}$ for $0\leq i \leq \frac{x-1}{2}$. 
	For $\frac{x+1}{2} \leq i \leq \frac{x+y}{2}-1$, let $K_i$ be the $2$-cycle $u_xv_{2i-x-1}v_{2i-x}v_{2i-x+1}$.
	Further let $K_{(x+y)/2}$ be $u_{x-1}u_xv_yv_{y-1}$.

	Note that every edge of $K$ is in exactly one of $K_1, \dots, K_r$.
	Further, every edge of $K_i$, $1 \le i \le r$, in the interior of~$K$ is in exactly two of $K_1, \dots, K_r$.
	Let $M_0, \dots, M_r$ be matchings such that the symmetric difference of $M_{i-1}$ and $M_i$ is $K_i$ for $i=1, \dots r$.
	Then by \cref{lemma:oneleggedcat_inside}, $M_{i-1}$ and $M_i$ are disjoint $\cater_3$-compatible.
	Further, the symmetric difference of $M_0$ and $M_r$ is $K$, implying that $M$ and $M'$ are connected in $\dcg(\cater_3)$. 
	
	Combining this result with the proof of \cref{caterpillar-compatible}, it follows that $\dcg(\cater_3)$ is connected.

	The bound on the diameter then follows from the bound on diameter of $\dcg(\tree)$ in combination with the fact that
	any set of disjoint inside cycles can be split into $\O(n)$ disjoint inside $2$-cycles.
\end{proof}

\section{Disjoint path-compatible matchings}\label{sec:path}

Let $\pa$ be the family of all spanning paths on $S$.
Note that paths are special instances of trees and caterpillars.
The following proposition states that in contrast to trees and caterpillars, $\dcg(\pa)$ is disconnected.

\begin{restatable}{proposition}{propPathThreeEars}\label{prop:Path:ThreeEars}
	Let $M$ be a plane matching on $S$ with at least three semiears. Then there is no spanning path on $S$ which is disjoint compatible to $M$, that is, $M$ is an isolated vertex in $\dcg(\pa)$.
\end{restatable}

	\begin{figure}[htb]
	\centering
	\vspace{-3ex}
	\includegraphics[page=1,scale=1]{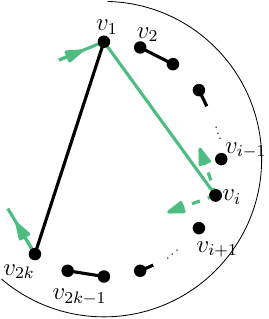}
	\caption{A $k$-semiear in a matching with a possibly disjoint \mbox{compatible} path drawn in green entering the ear at $v_1$ and leaving it at $v_{2k}$. If the path reaches vertex $v_i$, it can only traverse vertices with index either smaller or larger than $i$ afterwards.}
	\label{fig:3ears}
\end{figure}

\begin{proof}
	We claim that any semiear in a matching $M$ has to contain an end of a disjoint compatible path. By contradiction, we assume the contrary. We consider a $k$-semiear and label the $2k$ vertices of the semiear along the boundary of the convex hull by $v_i$, that is, the path enters the semiear at vertex $v_1$ and leaves it at $v_{2k}$.
	
	Observe that once the path reaches vertex $v_i$, it can only visit either vertices with smaller or larger index (cf. \ \cref{fig:3ears}). Therefore we need to go along them in ascending order. However this is not possible since there is an edge $v_2v_3$ and we demand disjoint compatibility. 
	This proves that in every semiear of $M$ any disjoint compatible path has to start or end there, thus no matching with three or more semiears contains such a path.
\end{proof}

From  \cref{prop:Path:ThreeEars} it follows that $\dcg(\pa)$ contains isolated vertices if $S$ is a set of at least $12$ points.
Note that there are also matchings with two semiears that are not compatible to any spanning path.
On the other hand, one might ask whether all matchings which are disjoint $\pa$-compatible to some other matching are in one connected component of $\dcg(\pa)$.
The following proposition gives a negative answer to that question.

\begin{restatable}{proposition}{propPerimeterDisconnected}\label{per_not_conn}
	The two perimeter matchings are not connected in $\dcg(\pa)$.
\end{restatable}

\begin{proof}
	\Wlog consider the even perimeter matching.
	We claim that every matching that is in the component of $\dcg(\pa)$ containing the even perimeter matching has only even semiears\footnote{Even (odd) semiears have only even (odd) perimeter edges.}.
	Assume to the contrary that there exist two disjoint $\pa$-compatible matchings $M$ and $M'$ such that $M$ has only even semiears and $M'$ has an odd semiear.
	Let $X_1$ and $X_2$ be two even semiears of $M$ and let $X'$ be an odd semiear of $M'$.
	Let  $D=X_1 \cup X_2 \cup X'$ be the union of the three semiears.
	We have four cases.
	
	\emph{Case 1.} The interior of $\Conv(X_1)$, $\Conv(X_2)$ and $\Conv(X')$ are disjoint.
	
	If $X_1 \cap X'=\emptyset$ and $X_2 \cap X' = \emptyset$, then $D$ is a plane matching with three ears.
	Hence, we have a contradiction by \cref{prop:Path:ThreeEars}.
	\smallskip
	
	\emph{Case 2.} $\Conv(X')$ is in $\Conv(X_1)$ or in $\Conv(X_2)$.  
	
	If $\Conv(X')$ is in $\Conv(X_1)$ or in $\Conv(X_2)$, then $D$ contains the boundary $\overline{X'}$ of $\Conv(X')$ since $X'$ only has odd perimeter edges and $X_1, X_2$ only have even perimeter edges.
	Hence, $D$ contains an ear which gives a contradiction to the assumption by~\cref{earandboundaryarea}.
	\smallskip
	
	\emph{Case 3.} $\Conv(X_1)$ is in $\Conv(X')$ or $\Conv(X_2)$ is in $\Conv(X')$.
	
	This case gives a contradiction in a similar way as Case 2.
	\smallskip
	
	\emph{Case 4.} The interiors of $\Conv(X')$ and $\Conv(X_1)$ intersect or the interiors of $\Conv(X')$ and $\Conv(X_2)$ intersect.
	
	\Wlog $\Conv(X')$ and $\Conv(X_1)$ intersect.
	Note that $X'$ and $X_1$ do not have any edge in common.
	Let $B$ be the boundary of $\Conv(X')\cap\Conv(X_1)$.
	Let $S_B$ be the points of $S$ that are in $B$.
	Every point of $S_B$ is incident to two edges of $D$ since $X'$ and $X_1$ are semiears and they do not have any edge in common.
	This means that every point of $S_B$ is incident to two perimeter edges except the two point that are incident to a diagonal.
	If $S_B$ contains at least three points, then $B$ is a boundary area which gives a contradiction to the assumption by~\cref{earandboundaryarea}.
	If $S_B$ contains only two points, then the perimeter edge of $X'$, that is incident to a point of $S_B$, and the perimeter edge of $X_1$, that is incident to a point of $S_B$, are both odd perimeter edges or are both even perimeter edges.
	Hence, $X'$ and $X_1$ are both odd semiears or even semiears, which is impossible since $X_1$ is an even semiear and $X'$ is an odd semiear.
	\smallskip
	
	This means that a component of $\dcg(\pa)$ containing a matching with only even semiears, does not contain a matching which has an add semiear.
	So the even perimeter matching is in a component of $\dcg(\pa)$ containing only matchings with only even semiears, while the odd perimeter matching is in a component of $\dcg(\pa)$ containing only matching with only even semiears.
\end{proof}

We remark that several more observations on $\dcg(\pa)$ can be found in~\cite{ObmannThesis}.

\section{Conclusion and discussion}

We have shown that the diameter of the disjoint $\tree$-compatible graph $\dcg(\tree)$
for point sets $S$ of $2n$ points in convex position is 4 or 5 when $2n\ge 10$.

We conjecture that the diameter of $\dcg(\tree)$ is $4$ for all $2n\geq 18$.
An open question is the computational complexity of determining whether two given matchings have distance $3$ in $\dcg(\tree)$.

For $\dcg(\cater)$ and $\dcg(\cater_3)$, we showed that their diameters are both in $\O(n)$.
Determining whether those two diameters are (asymptotically) the same, and what their precise values are, remains open.

Regarding spanning paths we showed that $\dcg(\pa)$ is disconnected, with no connection between the two perimeter matchings and many isolated vertices.

Further natural open questions include determining whether $\dcg(\tree)$ is connected for general point sets, and whether there exist point sets $S$ such that $\dcg(\pa)$ is connected.

We remark that our main approach for bounding diameters was to rotate inside semicycles.
A similar approach has also been used in a different setting of flip graphs of matchings.
A difference is that in that flip graph setting, semiears can be flipped, which is not possible in the disjoint $\tree$-compatible setting.
On the other hand, one can flip only one semicycle, or even only two edges at a time.
A recent related work on flip graphs
is \cite{Milich2021OnFI}.
There, so-called \emph{centered} flips in matchings on convex point sets are considered.
A centered flip is the rotation of an empty quadrilateral that contains the center of the point set.
This operation is more restrictive than our rotation of quadrilaterals for $\dcg(\cater_3)$,
as can also be seen by the fact that the flip graph of matchings with centered flips is sometimes disconnected.

\bibliographystyle{plain}
\bibliography{mtm_fullversion}

\end{document}